\documentclass[useAMS,usenatbib]{mn2e}
\usepackage{psfig}

\newcommand{\be}{\begin{equation}}
\newcommand{\en}{\end{equation}}

\def\kms{km\,s$^{-1}$~}

\title[HI absorption towards 3C\,452]{Associated 21-cm H{\sc i} absorption towards the  
radio galaxy 3C\,452 (J2245+3941)}
\author[Neeraj Gupta and D.J. Saikia]{Neeraj Gupta\thanks{E-mail:neeraj@ncra.tifr.res.in (NG); 
djs@ncra.tifr.res.in (DJS)} and D.J. Saikia \\
National Centre for Radio Astrophysics, TIFR, Pune 411 007, India \\
} 
\begin{document}

\date{Accepted. Received; in original form }

\pagerange{\pageref{firstpage}--\pageref{lastpage}} \pubyear{2005}

\maketitle

\label{firstpage}

\begin{abstract}
We report the detection of 21-cm H{\sc i} absorption towards the core of the Fanaroff-Riley\,II 
radio galaxy 3C\,452 (J2245+3941). 
The absorption profile is well resolved into three components;  the strongest and 
narrowest component being coincident with the velocity corresponding to [O~{\sc iii}] 
emission lines while the other two components are blue-shifted with respect to it 
by $\sim$30 amd $\sim$115 \kms.  If the systemic velocity of the host galaxy is 
determined from low-ionization lines, which are red-shifted with respect to the [O~{\sc iii}] 
doublet by about $\sim$200 km\,s$^{-1}$, then both the [O~{\sc iii}] emission 
and 21-cm absorption lines are associated with outflowing material.  
The neutral hydrogen column density is estimated to be 
N(HI)=6.39$\times$10$^{20}$(T$_s$/100)($f_c$/1.0)$^{-1}$ cm$^{-2}$, where T$_s$ and $f_c$ 
are the spin temperature and partial coverage of the background source respectively. 
If the 21-cm absorber is also responsible for the nuclear extinction at infrared
wavelengths and x-ray absorption, then for a spin temperature of $\sim$8000 K, 
the absorber occults only $\sim$10 per cent of the radio core. 
\end{abstract}

\begin{figure*}
\centerline{\vbox{
\psfig{figure=bw_map.ps,width=19cm,height=14cm,angle=-90}
\vskip -6.5cm
\hskip -4.0cm
\psfig{figure=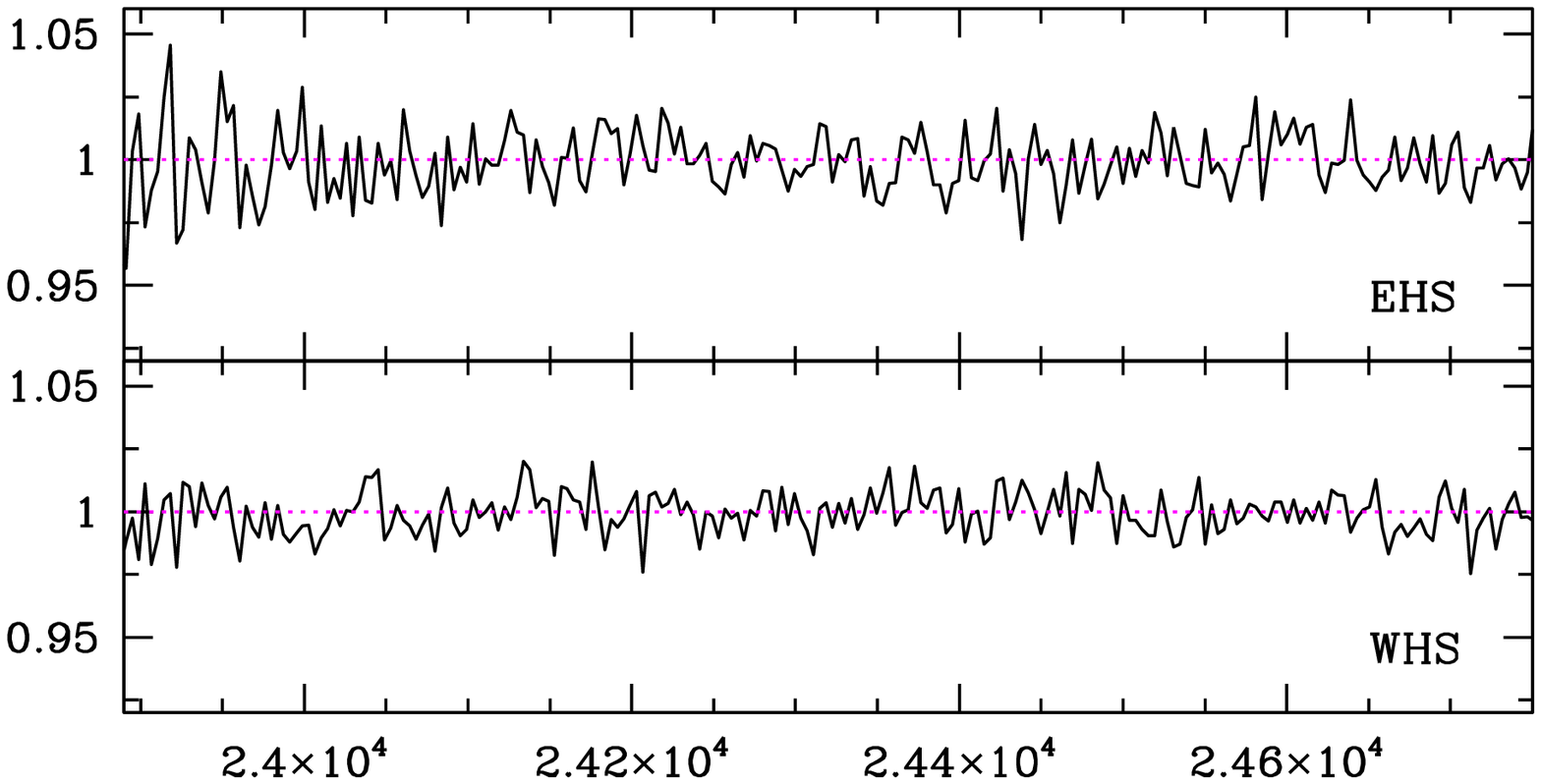,width=8cm,height=6.5cm,angle=0}
\vskip -16.0cm
\hskip +8.0cm
\psfig{figure=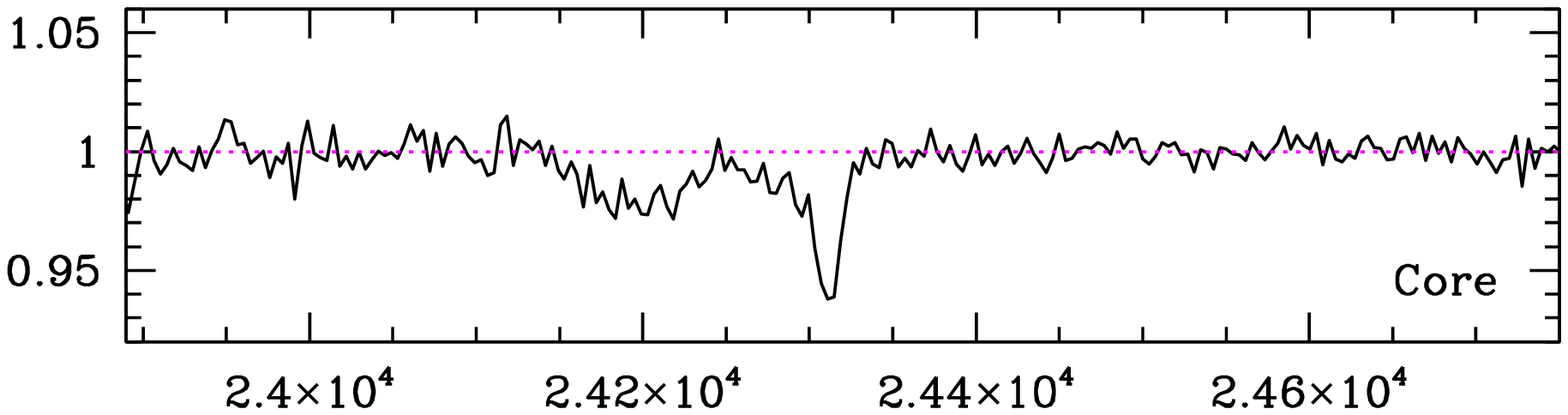,width=9.5cm,height=10cm,angle=0}
}}
\vskip +6.0cm
\caption[]{ GMRT image of 3C\,452 with an rms noise of 0.3 mJy/beam.  The contour levels are 
1.5$\times$($-$2,$-$1, 1, 4, 8 , 16, 20, 32, 64, 128) mJy/beam.  The restoring beam of 
3.14$^{\prime\prime}\times$2.25$^{\prime\prime}$ along position angle 52$^\circ$ is shown 
as an ellipse and the position of the optical host galaxy is marked with a cross. 
The spectra at the peak intensity pixel in the eastern hotspot (EHS), 
core and western hotspot (WHS) are also shown.  For these x-axis and y-axis are 
heliocentric velocity in km s$^{-1}$ and normalised intensity respectively.
}
\vskip -12.1cm
\hskip +4.5cm
\begin{picture}(200,200)(-100,-100)
\put(-45,10){\vector(1,1){80}}
\put(-51,3){\Large +}
\end{picture}
\vskip -5.9cm
\hskip 9.0cm
\begin{picture}(200,200)(-100,-100)
\put(43,18){\vector(-3,-2){130}}
\end{picture}
\vskip -8.0cm
\hskip +2.7cm
\begin{picture}(200,200)(-100,-100)
\put(-169,47){\vector(-1,-3){30}}
\end{picture}
\vskip +4.5cm
\label{map}
\end{figure*}

%

\section{Introduction}
An understanding of the properties of the gaseous environments of radio galaxies 
and quasars could provide valuable insights towards understanding the phenomenon 
of radio activity associated with these objects and their evolution. Such studies
also enable us to test consistency of these properties with the unified schemes
for these objects. An important way of probing the neutral component of this gas
over a wide range of length scales is via 21-cm H{\sc i} absorption towards 
radio sources of different sizes. These range from the sub-galactic sized compact 
steep-spectrum (CSS) and gigahertz peaked-spectrum (GPS) sources, which are believed
to be young ($<$10$^5$ yr), as compared with the larger sources which could extend 
to over a Mpc and are typically $\sim$10$^8$ yr old.
 
H{\sc i} absorption lines are seen more often towards CSS and GPS objects, with 
the H{\sc i} column densities being anticorrelated with the source sizes. The 
absorption profiles exhibit a variety of line profiles, suggesting complex gas motions
(Vermeulen et al.  2003; Pihlstr\"om, Conway \& Vermeulen 2003; Gupta et al. 2006). 
Gupta \& Saikia (2006) have examined whether the H{\sc i} column density is consistent 
with the unified scheme for radio galaxies and quasars by using core prominence as an 
indicator of the orientation of the jet axis to the line of sight for a sample of 32 
CSS and GPS objects. They find the 
relationship between the H{\sc i} column density and core prominence to be consistent 
with the unified scheme in a paradigm where the  the H{\sc i} gas is distributed in 
a circumnuclear disk with a scale smaller than the size of the compact radio sources.  
Very Long Baseline Interferometry (VLBI)-scale spectroscopic observations towards 
several sources such as the CSS object
J0119+3210 (4C+31.04) and the cores of  
larger sources such as the well known Fanaroff-Riley class\,II (FR\,II) source
Cyg\,A and the FRI objects NGC\,4621 and Hydra\,A
show evidence of absorption arising from a circumnuclear disk-like structure 
(Conway \& Blanco 1995; Taylor 1996; Conway 1999; van Langevelde et al. 2000).  
Morganti et al. (2001) have suggested that there may be differences in the torus/disk
between FR\,I and FR\,II sources, with those in FR\,Is being geometrically thin.

Compared with 21-cm absorption searches towards CSS and GPS sources the extended 
radio galaxies have received relatively less attention.  
Notable exceptions to this are the studies by van Gorkom et al. (1989) and 
Morganti et al. (2001) whose samples also contained larger radio sources in addition to
the subgalactic-sized compact ones.   
To investigate whether the circumnuclear gas evolves 
as the source grows from CSS and GPS scales to the large radio sources, one needs to 
probe the distribution and properties of 
this gas on similar scales with comparable sensitivities. 
This can be achieved via 21-cm absorption measurements towards the
compact cores of the large objects using high-resolution observations where the core is
clearly resolved from the more extended bridge emission. Such observations will also
help clarify any differences between FR\,I and FR\,II sources. However,  
very few such measurements exist in the literature. Amongst large FR\,II radio galaxies,
two notable detections are Cyg\,A and 3C\,353 (Conway \& Blanco 1995; Morganti et al. 2001).
In this paper, we report the discovery of 21-cm 
absorption towards the core of the FR\,II radio galaxy 3C\,452 with the Giant
Metrewave Radio Telescope (GMRT), and examine the properties of 
this 21-cm absorber in the light of observations at other wavelengths.   

%
%
\begin{figure}
\centerline{\vbox{
\psfig{figure=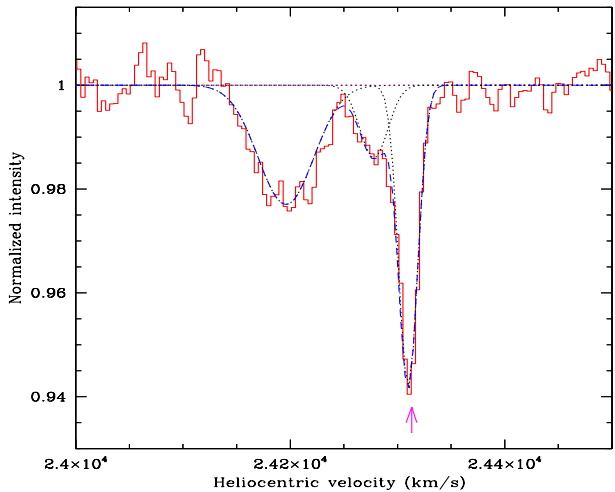,height=7.0cm,width=8.5cm,angle=0}
}}
\caption[]{ The H{\sc i} absorption spectrum (histogram) towards the core of the radio galaxy 3C\,452.  
The spectrum has been smoothed using a 3-pixel wide boxcar filter.  The three Gaussian components 
fitted to the absorption profile and the sum of these components i.e. the fit are plotted as 
dotted and dashed lines respectively.  The systemic velocity (24313 km s$^{-1}$) determined using 
[O~{\sc iii}] emission lines has been marked by an arrow.
} 
\label{fits}
\end{figure}

\section{3C\,452 (J2245+3941)}
3C\,452 is a high-excitation narrow-line FR\,II radio galaxy at the 
redshift, $z_{em}$=0.08110 determined using the [O~{\sc iii}]~$\lambda\lambda$4959, 5007\,
\AA~ emission lines (Lawrence et al. 1996; Jackson \& Rawlings 1997).  
This is consistent with the estimate from the H$\beta$ line within $\sim$30 km\,s$^{-1}$.  
However from the data of Lawrence et al. we find that the low-ionization forbidden emission 
lines namely, 
[O~{\sc i}], [O~{\sc ii}], [N {\sc ii}] and [S~{\sc ii}] are all systematically 
red-shifted with respect to it by $\sim$200 km\,s$^{-1}$.  Further, the low-ionization lines 
appear broader (FWHM$\sim$900 km\,s$^{-1}$) than the high-ionization 
ones which have a FWHM of $\sim$600 km\,s$^{-1}$, suggesting that the low-ionization lines
may be arising from shock-ionised gas (see Villar-Martin et al. 1999).  
The total radio luminosity of 3C\,452 at  
5 GHz is 3.6$\times$10$^{25}$ W Hz$^{-1}$ sr$^{-1}$ (H$_o$=71 km s$^{-1}$ Mpc$^{-1}$, $\Omega_m$=0.27, 
$\Omega_\Lambda$=0.73; Spergel et al. 2003), which is well above the dividing
line for the two FR types. At radio wavelengths the source exhibits a 
symmetric triple morphology with a largest angular size of 256 arcsec, which 
corresponds to a linear size of 386 kpc (e.g. Black et al. 1992; 
Dennett-Thorpe et al. 1999). The compact arcsec-scale radio core is resolved
with an angular resolution of $\sim$1.2 mas into a central peak of emission,
which we call the `nucleus', and symmetric jet-like structures on opposite sides
of it (Giovannini et al. 2001). Most of the emission is within $\sim$5 mas (7.6 pc)
of the radio nucleus, although faint emission is seen extending upto $\sim$20 mas 
(30 pc) from the radio nucleus on both sides.  
Using the jet symmetry and core prominance at 5 GHz, Giovannini et al. suggest that 
this source must be oriented at an angle greater than $\sim$60$^\circ$  to the
line of sight. 
The core as well as diffuse emission associated with the lobes have been 
detected in x-rays using the Chandra telescope (Isobe et al. 2002).  
The HST WFPC-2 image of the host galaxy suggests the presence of a faint dust lane 
roughly perpendicular to the radio axis of the source (de Koff et al. 2000).   

%
\section{Observations and data analyses}
We observed 3C\,452 with the GMRT to search for 
associated 21-cm absorption towards the radio source.  3C\,452 was observed on 2005 Dec 11 
with a bandwidth of 4 MHz for $\sim$9 hours.  
In these observations  we made use of the new high-resolution mode of the 
GMRT correlator software. This allows one to split the baseband bandwidth into 256 channels,
instead of the usual 128 channels.  This provided us a 
spectral resolution of $\sim$3.6 \kms.
The local oscillator chain and FX correlator 
system were tuned to centre the baseband bandwidth at 1313.85 MHz, 
the redshifted 21-cm frequency corresponding to $z_{em}$=0.0811.  We observed standard 
flux density calibrators 3C\,286 and 3C\,48 every 3 hours to correct for the variations 
in amplitudes 
and bandpass.  The compact radio source J2202+4216 was also observed approximately 
every 35 minutes for phase calibration of the array.  A total of $\sim$5.6 hours of data on source 
were acquired in both the circular polarization channels RR and LL. 

The data were reduced in the standard way using Astronomical Image Processing System (AIPS) 
package.  After the initial flagging or editing of bad data and calibration, source and 
calibrator data were examined for baselines and timestamps affected by Radio Frequency 
Interference (RFI).  These data were excluded from further analysis.  A continuum image 
of the source was made using calibrated data averaged over 60 line-free channels.  
Due to our interest in absorption towards compact components like the core, the imaging 
was done without any {\it uv} cut-off or tapering in the visibility plane. This provided us 
with the highest possible resolution.   
This image was then self-calibrated until a satisfactory map was obtained (see Fig.~\ref{map}).   
The self-calibration complex gains determined from this were applied to all the 256 
frequency channels and continuum emission was subtracted from this visibility data cube using 
the same map.  Spectra at various positions on the radio galaxy 3C\,452 were extracted from 
this cube. The whole process was done separately for Stokes RR and LL to check for consistency.  
The two polarization channels were then combined to get the final Stokes I spectrum, which 
was then shifted to the heliocentric frame. 
    
\section{Results and discussion}
%
\begin{table} 
\caption{Observational results} 
\begin{center}
\begin{tabular}{lccrr} 
\hline 
Component   &   Peak flux  & $\sigma$ & $\tau_{1\sigma}$ & N(H{\sc i})       \\
            &     Jy/b     &  mJy/b   &    10$^{-3}$     & 10$^{19}$cm$^{-2}$\\
\hline 
EHS         &  0.094     &  1.1     & $<$11.7          &   $<$13.6    \\
Core$^\ast$ &  0.194     &  1.0     &     $-$          &       $-$    \\
WHS         &  0.128     &  1.1     & $<$8.59          &   $<$9.95    \\
\hline 
\end{tabular}
\end{center}
Col. 1: radio component; col. 2: peak flux density; 
col. 3: rms noise in the spectrum; col. 4: optical depth estimate and 
col. 5: H{\sc i} column density or 3$\sigma$ upper limit to it.  
T$_s$=100 K and $\Delta v$=20 \kms have been adopted for estimating N(H{\sc i}) 
upper limits. \\
$^\ast$  See Table~\ref{gauss}. 
\label{results} 
\end{table} 
%
\begin{table}
\caption{Multiple Gaussian fit to the H{\sc i} absorption spectrum}
\begin{center}
\begin{tabular}{|c|l|l|c|c|}
\hline
Id. &   v$_{\rm{hel}}$ &  FWHM & Frac. abs. & N(H{\sc i}) \\
no. &               &               & & 10$^{20}$($\frac{T_s}{100})(\frac{f_c}{1.0})^{-1}$  \\
    &   km s$^{-1}$ &   km s$^{-1}$ & & cm$^{-2}$ \\
\hline
1 & 24196         & 61(3) & 0.023(0.001)&   2.71    \\
2 & 24278         & 31(4) & 0.014(0.002)&   0.84    \\   
3 & 24310         & 21(1) & 0.058(0.002)&   2.35    \\  
\hline
\end{tabular}
\end{center}
\label{gauss}
\end{table}
%
%
In Fig.~\ref{map} we present our GMRT image of 3C\,452 at 1313 MHz, which
shows the radio core, the symmetrically located hotspots on opposite sides
of the parent optical galaxy and the extended bridge of emission. The peak
flux density of the radio core in our image is 194 mJy/beam.   
The peak intensities for the brightest components in the eastern and western lobes, 
referred to as the eastern hotspot (EHS) and the western hotspot (WHS) 
are given in Table.~\ref{results}.  
The total flux density in the image is 9.98 Jy at 1313 MHz, which is within a few per cent
of the value of 10.2 Jy.  The latter has been estimated using the measurements with least
errors at 750 and 4850 MHz from the NASA Extragalactic Database.
The H{\sc i} absorption spectra towards the core and hotspots of 3C\,452 are 
presented in Fig.~\ref{map}.  H{\sc i} absorption has been detected clearly
towards the core of the radio galaxy while no absorption has been detected towards 
the hotspots or any other part of the source. The rms noise in the spectra and 
1-$\sigma$ upper limits to the optical depth towards EHS and WHS are presented 
in Table.~\ref{results}. The H{\sc i} column density, N(H{\sc i}), integrated over the
entire spectrum using the relation
\begin{equation}
{\rm N(HI)}=1.835\times10^{18}\frac{{\rm T}_{s}~\int{\tau(v)dv}}{f_c}~ {\rm cm^{-2}},
\label{eq1}
\end{equation}
where T$_s$, $\tau$ and $f_c$ are the spin temperature, optical depth at a velocity
$v$ and the fraction of background emission covered by the absorber respectively,
is 6.39$\times$10$^{20}$(T$_s$/100)(1.0/$f_c$) cm$^{-2}$.  
The absorption profile is 
well fitted with three Gaussian components (Fig.~\ref{fits}).  The best fit parameters 
are summarised in Table~\ref{gauss}.  
It may be noted that the widths of the components and corresponding column densities are typical 
of those observed in our Galaxy (e.g. Ferri\`ere 2001; Dickey \& Lockman 1990).  
The absorption profile consists of a strong narrow component 
near the velocity of the galaxy estimated from the [O~{\sc iii}] emission lines 
and two broader components blue-shifted with respect to it by about $\sim$30 and 
$\sim$115 \kms. However different velocities for the host galaxy estimated from different
emission lines can sometimes make it difficult to ascertain the kinematics of the
absorbing gas (see e.g. Tadhunter et al. 2001; Vermeulen et al. 2006). In 3C\,452 the 
velocity corresponding to the low-ionization
lines would imply that the narrow and two broad 21-cm absorption lines
are blue-shifted by $\sim$200, 230 and 315 km\,s$^{-1}$. This would then imply that
both the [O~{\sc iii}] emission and 21-cm absorption lines are associated with outflowing gas.
Fast and broad outflows in H{\sc i} with velocities extending up to
several thousand km s$^{-1}$ have been reported for a number of radio galaxies
(Morganti, Tadhunter \& Oosterloo 2005).
%

Amongst the two notable detections of 21-cm absorption towards the cores of extended 
FR\,II radio galaxies, namely Cyg\,A and 3C\,353, 
the system associated with the narrow-line radio galaxy Cyg\,A has a 
FWHM of $\sim$270 km\,s$^{-1}$ and column density of 
2.54$\times$10$^{21}$(T$_s$/100)(1.0/$f_c$) cm$^{-2}$ (Conway \& Blanco 1995).    
The redshift of Cyg\,A determined from the [O~{\sc iii}] emission lines is $z_{em}$=0.05606 
(see Section 4.3 of Tadhunter et al. 2003).  The 21-cm absorption line associated with 
it consists of two components; the narrower (FWHM$\sim$102 km\,s$^{-1}$) and weaker one being 
consistent with the $z_{em}$ while the other one (FWHM$\sim$151 km\,s$^{-1}$) is red-shifted with 
respect to it by $\sim$170 km\,s$^{-1}$. 
In the weak-line radio galaxy 3C\,353, the absorption-line (FWHM$\sim$300 km\,s$^{-1}$) 
corresponds to a total H{\sc i} column density of 
4.2$\times$10$^{21}$(T$_s$/100)(1.0/$f_c$) cm$^{-2}$ (Morganti et al. 2001).  
The peak of absorption is blue-shifted with respect to the systemic velocity of the 
host galaxy estimated from the [O~{\sc iii}] and H$\alpha$ emission lines by $\sim$100 km\,s$^{-1}$.  
Thus 21-cm absorption-line systems detected towards the cores of Cyg\,A, 3C\,353 and 3C\,452 
have peaks of absorption red-shifted, blue-shifted and consistent with respect to the velocity 
corresponding to the [O~{\sc iii}] emission line.  
Thus the absorption profiles are often complex suggesting radial motions (i.e infall or outflow) 
in addition to disks, as has been inferred for the CSS and GPS objects.
%

It is interesting to note that the low-ionization forbidden emission lines in 3C\,452 are 
found to be broader and more red-shifted than the high-ionization ones.  
The low-ionization lines could then correspond to gas interacting with the jet, 
thereby being broadened.  
In that case, a plausible scenario to explain the different 21-cm absorption components 
is one in which 
[O~{\sc iii}] emission lines represent the systemic velocity and the narrowest 21-cm 
absorption line component coincident with it, arises from gas in a circumnuclear disk 
aligned with the 
dust lane perpendicular to the radio source axis of 3C\,452 (de Koff et al. 2000).  
The possibility of absorption occurring in such disks is consistent with
a number of observational trends. For example, Morganti et al. (2001) detect
21-cm absorption towards 3 out of 4 narrow-line radio galaxies, while no 
absorption was detected towards 4 broad-line radio galaxies. The latter are
expected to be inclined at smaller angles to the line of sight so that the
disk is unlikely to produce significant H{\sc i} absorption. These results
of Morganti et al. are also consistent with the higher detection rate of 
21-cm H{\sc i} absorption towards radio galaxies compared with quasars (Vermeulen et 
al. 2003; Gupta et al. 2006). 
Gupta \&  Saikia (2006) for a sample of CSS and GPS sources find the dependence 
of  N(H{\sc i}) on the degree of core prominence to be consistent with the H{\sc i} 
having a disk-like distribution, with the source sizes being larger than the scale
size of the disk. Thus, if we attribute the narrow component to be due to
absorption by a disk, the broader blue-shifted components could still be due to 
outflowing gas. 

%
In the following we compare our results with observations at infrared and x-ray
wavelengths to constrain some of the properties of the absorbing material.      
The nuclear extinction of 3C\,452 as measured by near-infrared observations is 
A$_{\rm v}>$2.30 (Marchesini, Capetti \& Celotti 2005).  For the diffuse ISM of our 
Galaxy, the mean ratio of total neutral hydrogen i.e. N(H{\sc i}+H$_2$) is related to 
visual extinction as
\begin{equation}
{\rm N(HI+H_2)=1.89\times10^{21}A_v \,mag^{-1}cm^{-2} } 
\label{extinction}
\end{equation}
(Bohlin, Savage \& Drake 1978; Cardelli, Clayton \& Mathis 1989).  
Using this relation and assuming that all the gas producing extinction is atomic we 
get for 3C\,452, N(H{\sc i})$>$2.2$\times$10$^{21}$ cm$^{-2}$. Its comparison with the 
total neutral hydrogen column density determined from 21-cm absorption 
implies T$_s>$350 K for $f_c$=1.0.  In fact, the spin temperature of 21-cm absorbing 
gas can be much higher than $\sim$350 K.  For the warm neutral medium seen in the Galaxy 
T$_s$ ranges from 5000$-$8000 K (Kulkarni \& Heiles 1988).  
Such high spin temperatures are also expected to arise in the proximity of the active
nucleus (Bahcall \& Ekers 1969).    

%
Using the Chandra telescope, Isobe et al. (2002) detected x-ray emission associated with 
the core and lobes of the radio galaxy.  They infer that a total hydrogen column 
density N(H)$\approx$6$\times$10$^{23}$ cm$^{-2}$ is required to fit the x-ray 
spectrum towards the core.  
Even for a spin temperature of $\sim$8000 K, the 21-cm absorbing gas has atleast an order 
of magnitude lesser column density than that required for x-ray absorption. This either
implies that the x-ray  absorbing gas is different from that responsible for 21-cm
absorption or that the 21-cm absorber does not cover the core completely. 
The nucleus in the 5-GHz VLBI map of Giovannini et al. (2001) 
has a flux density of only $\sim$20 mJy.  Assuming a spectral index of 0, this would then 
correspond to 
$f_c\sim$0.1.  Thus a spin temperature of $\sim$8000 K and $f_c$ of $\sim$0.1 are 
required if the gas producing 21-cm absorption and nuclear extinction also causes the 
x-ray absorption.  
It is worth mentioning here that reconciliation of x-ray absorbing column density 
with the absorbers detected at other wavelengths in AGN is a recurring problem 
(see e.g. Hamann 1997; Gallimore et al. 1999).  
In the simplest explanation the x-ray absorber and the UV or 21-cm absorption 
line systems are not cospatial. For example, in a 21-cm absorption study of a sample of 13 Seyfert 
galaxies, Gallimore et al. (1999) did not find any correlation between the 21-cm and x-ray absorbing 
column density.  The nuclear extinction measured for a complete sub-sample of 3CR radio 
sources, which includes 3C\,452, ranges from 0 to 9 mag (Marchesini et al. 2004).  
Adopting 9 mag as an upper limit on A$_{\rm v}$ for 3C\,452 and $f_c$=1.0 for 
the 21-cm absorber would then imply that the 
gas producing 21-cm absorption and nuclear extinction, and x-ray absorption are not cospatial. 
H{\sc i} observations via VLBI techniques 
are required to constrain the properties and location of the absorbing clouds and 
further investigate this scenario.
The x-ray spectrum of diffuse emission associated with the lobes is consistent with the 
Galactic value of N(H)=1.2$\times$10$^{21}$ cm$^{-2}$.  This is consistent with the limits 
on the neutral hydrogen column density derived from the GMRT spectra towards the 
radio hotspots (Table~\ref{results}).

\section {Summary}
We have reported the detection of 21-cm absorption towards the core of the FR\,II radio 
galaxy 3C\,452.  The absorption profile is resolved into three components.  
The deepest and narrowest of these is consistent with the velocity 
corresponding to [O~{\sc iii}] emission lines.  The other two broader components 
are blue-shifted with respect to it by $\sim$30 and $\sim$115 \kms.  
If the systemic velocity of the host galaxy is determined from low-ionization forbidden 
emission lines then both the [O~{\sc iii}] emission and 21-cm absorption lines are 
associated with outflowing material. The 21-cm components are blue-shifted relative 
to the low-ionization lines by  $\sim$200, 230 and 315 \kms respectively.
The neutral hydrogen column density of the gas is estimated to be 
N(HI)=6.39$\times$10$^{20}$(T$_s$/100)($f_c$/1.0)$^{-1}$ cm$^{-2}$.  
If the 21-cm absorber is also responsible for the nuclear absorption seen at
x-ray wavelengths, then for a spin temperature of $\sim$8000 K, the absorber
occults only $\sim$10 per cent of the radio core. This would also be
consistent with the nuclear extinction seen at infrared wavelengths.
In our GMRT spectrum we do not detect 21-cm absorption towards locations other than 
the radio core of 3C\,452.  The upper limit on N(H{\sc i}) derived from spectra 
towards the hotspots is consistent with constraints on N(H) obtained from the 
diffuse x-ray emission associated with the radio lobes.
%
%
\section*{Acknowledgments}
We thank Raffaella Morganti, the referee, and Raghunathan Srianand for many useful suggestions and comments. 
We also thank the numerous contributers to the GNU/Linux group. This research has made use of 
the NASA/IPAC Extragalactic Database (NED) which is operated by the Jet Propulsion Laboratory, 
California Institute of Technology, under contract with the National Aeronautics and Space 
Administration. We thank the staff of GMRT for their assistance during our observations.  
The GMRT is a national facility operated by the National Centre for Radio Astrophysics of the Tata 
Institute of Fundamental Research. 

%

\label{lastpage}

\end{document}